\documentclass[journal]{IEEEtran}
\usepackage{amsthm,amsmath,amssymb}
\usepackage{eucal}
\usepackage{xifthen}
\usepackage{mathtools}
\usepackage{enumerate}
\usepackage{microtype}
\usepackage{xspace}
\usepackage{bm}
\usepackage{cite}
\usepackage{fancyhdr}
\usepackage{lastpage}
\usepackage[small]{caption}
\usepackage{xcolor}
\allowdisplaybreaks

\long\def\comment#1{}

\newfont{\bbb}{msbm10 scaled 700}

\newfont{\bb}{msbm10 scaled 1100}
\newcommand{\CC}{\mbox{\bb C}}

\newcommand{\mbs}[1]{\bm{#1}}
\newcommand{\vect}[1]{{\lowercase{\mbs{#1}}}}
\newcommand{\mat}[1]{{\uppercase{\mbs{#1}}}}

\newcommand{\T}{{\scriptscriptstyle\mathsf{T}}}

\renewcommand{\Re}[1][]{\ifthenelse{\isempty{#1}}{\operatorname{Re}}{\operatorname{Re}\left(#1\right)}}
\renewcommand{\Im}[1][]{\ifthenelse{\isempty{#1}}{\operatorname{Im}}{\operatorname{Im}\left(#1\right)}}

\newcommand{\av}{\vect{a}}

\newcommand{\hv}{\vect{h}}

\newcommand{\nv}{\vect{n}}

\newcommand{\pv}{\vect{p}}
\newcommand{\qv}{\vect{q}}

\newcommand{\sv}{\vect{s}}

\newcommand{\uv}{\vect{u}}
\newcommand{\vv}{\vect{v}}
\newcommand{\wv}{\vect{w}}
\newcommand{\xv}{\vect{x}}
\newcommand{\yv}{\vect{y}}

\newcommand{\phiv}{\hbox{\boldmath$\phi$}}

\newcommand{\Psim}{\hbox{\boldmath$\Psi$}}

\newcommand{\Am}{\mat{a}}

\newcommand{\Wm}{\mat{w}}
\newcommand{\Xm}{\mat{x}}

\newcommand{\IM}{\mat{i}}

\newcommand{\Ec}{{\mathcal E}}

\newcommand{\Gc}{{\mathcal G}}

\newcommand{\Kc}{{\mathcal K}}

\newcommand{\Mc}{{\mathcal M}}

\newcommand{\Xc}{{\mathcal X}}

\newcommand{\one}{\mat{\mathrm{1}}}
\newcommand{\zero}{\mat{\mathrm{0}}}

\newcommand{\CN}[1][]{\ifthenelse{\isempty{#1}}{\mathcal{N}_{\mathbb{C}}}{\mathcal{N}_{\mathbb{C}}\left(#1\right)}}
\renewcommand{\P}[1][]{\ifthenelse{\isempty{#1}}{\mathbb{P}}{\mathbb{P}\left(#1\right)}}
\newcommand{\E}[1][]{\ifthenelse{\isempty{#1}}{\mathbb{E}}{\mathbb{E}\left(#1\right)}}
\renewcommand{\det}[1][]{\ifthenelse{\isempty{#1}}{\mathrm{det}}{\mathrm{det}\left(#1\right)}}
\newcommand{\trace}[1][]{\ifthenelse{\isempty{#1}}{\mathrm{tr}}{\mathrm{tr}\left(#1\right)}}
\newcommand{\rank}[1][]{\ifthenelse{\isempty{#1}}{\mathrm{rank}}{\mathrm{rank}\left(#1\right)}}
\newcommand{\diag}[1][]{\ifthenelse{\isempty{#1}}{\mathrm{diag}}{\mathrm{diag}\left(#1\right)}}

\renewcommand{\arg}{{\hbox{arg}}}

\renewcommand{\Re}{{\rm Re}}
\renewcommand{\Im}{{\rm Im}}

\newcommand{\herm}{{\sf H}}

\allowdisplaybreaks

\DeclareMathAlphabet{\mathcal}{OMS}{cmsy}{m}{n}

\newtheorem{remark}{Remark}

 \usepackage{hyperref}
 \hypersetup{
     bookmarks=false,         
     unicode=false,          
     pdftoolbar=true,        
     pdfmenubar=true,        
     pdffitwindow=false,     
     pdfstartview={FitH},    
     pdfnewwindow=true,      
     colorlinks=true,       
     linkcolor=red,          
     citecolor=green,        
     filecolor=magenta,      
     urlcolor=cyan           
 }
\usepackage{subfigure}
\usepackage{cite}
\usepackage[final]{graphicx}
\usepackage{amsmath}
\usepackage{amssymb}
\usepackage{amsfonts}
\usepackage{array}
\usepackage{amsfonts}
\usepackage{stfloats}
\usepackage{lineno}
\usepackage{colortbl}
\usepackage{algorithmic}
\usepackage{epstopdf}
\usepackage{amsthm}
\usepackage{multirow}
\usepackage{multicol}
\usepackage{graphicx}
\usepackage{lipsum}
\usepackage[margin=0.7in]{geometry}
\newcolumntype{C}{>{\centering\arraybackslash$}p{\linewidth}<{$}}
\usepackage[linesnumbered,ruled,vlined,lined]{algorithm2e}
\theoremstyle{plain}
\theoremstyle{remark}
\usepackage{caption}
\usepackage{subfigure}

\allowdisplaybreaks  
\usepackage{tikz}
\newcommand{\tikzxmark}{%
\tikz[scale=0.2] {
    \draw[line width=0.8,line cap=round] (0,0) to [bend left=5] (1,1);
    \draw[line width=0.6,line cap=round] (0.1,0.85) to [bend right=3] (0.8,0.05);
}}
\begin{document}
\title{Topology-Aware Integrated Sensing, Communication, Charging in Massive Low-Altitude Wireless Network}

\author{Han Yu,~\IEEEmembership{Member, IEEE}, Jiajun He,~\IEEEmembership{Member, IEEE},~Zhaofeng Liu,~\IEEEmembership{Member, IEEE}, \\ 
and~Hing Cheung So,~\IEEEmembership{Fellow, IEEE}
\thanks{H. Yu is with the Faculty of Electrical Engineering and Computer Science, Technical University of Berlin, Germany (email: han.yu.1@campus.tu-berlin.de).}
\thanks{J. He is with the Department of Electrical Engineering (ISY), Linköping University, Linköping, SE-58183, Sweden. (email: frankho1366@gmail.com).}
\thanks{Z. Liu, Northeastern University, US (zh.liu@northeastern.edu).}
\thanks{H. C. So is with the Department of Electrical Engineering, City University of Hong Kong, Kowloon, Hong Kong SAR, China (email: hcso@ee.cityu.edu.hk).}
}
	
\maketitle

\begin{abstract}
    Future low-altitude wireless networks (LAWNs) are expected to simultaneously support sensing, communication, and charging, resulting in tightly coupled multi-objective optimization problems with strong interdependencies among heterogeneous functions. However, existing multi-objective frameworks typically rely on complex problem-specific formulations and alternating optimization procedures, which suffer from high computational complexity and limited scalability in large-scale, highly dynamic deployments. In this paper, we propose a unified topology-aware (TA) framework that abstracts terrestrial and non-terrestrial devices as nodes, and models their interactions as edges, forming a bipartite graph representation of the LAWN. By leveraging this graph representation, we develop a low-complexity topology reconfiguration algorithm based on a unified resource-adjustment rule. 
    Simulation results demonstrate that the proposed TA-based method consistently outperforms state-of-the-art approaches in multi-functional coordination efficiency, while maintaining robustness and scalability in scenarios involving massive terrestrial and airborne user equipments.
\end{abstract}


\section{Introduction}

Low-altitude wireless networks (LAWNs) have attracted significant attention in recent years, driving the development of a wide range of emerging applications, including aerial surveillance, environmental monitoring, and intelligent transportation systems \cite{11017717, 11315847, he2026LAWN}. In LAWNs, airborne platforms, such as unmanned aerial vehicles (UAVs), can be assigned heterogeneous functional roles to support a diverse set of network services. For instance, UAVs is able to perform environmental sensing to extract spatial and channel state information and construct channel knowledge maps (CKMs) \cite{CKM2026, 11320819}, which can subsequently be shared with terrestrial networks to support adaptive beamforming and dynamic resource allocation. In addition, UAVs can be deployed as aerial base stations (BSs) or mobile relays to extend network coverage in scenarios where terrestrial links suffer from severe blockage effects \cite{7572068, blockage}. Given the limited battery capacity of the miniature UAVs or drones, integrating wireless power transfer (WPT) into LAWNs is essential. In this context, both terrestrial and non-terrestrial BSs can function as WPT nodes to collaboratively provide energy resources to these UAVs \cite{9163290, 9206056}, thereby extending their operational lifetime. Nevertheless, supporting multiple functions within a LAWN, such as UAV communication, target sensing and localization, and WPT, is challenging. For example, in integrated sensing and communication (ISAC) scenarios, sensing operations may introduce significant interference that degrades communication performance, while the integration of multiple functionalities further increases the complexity of system coordination and performance optimization \cite{11155112, 11358925}. Therefore, there is a need to develop a scalable, efficient optimization framework for enabling multiple functions.

\vspace{-0.2cm}

\subsection{Related Work}

The diverse and dynamic functionalities of the LAWN underscore the significance of an efficient coordination scheme to fully unlock its potential. Jin $et$ $al.$ \cite{MO_SC2} consider the co-design of UAV signaling and target sensing in multi-drone cooperative systems, where the joint communication and sensing problem is formulated using a linear quadratic regulator (LQR) cost function and solved via an alternating optimization (AO) approach. Wu $et$ $al.$ \cite{Wu2026LAWN} propose an energy-efficient ISAC framework for the control and sensing of multiple UAV formations in LAWNs. The resulting optimization problem is also formulated as an LQR minimization problem, but is solved using successive convex approximation (SCA) and semidefinite relaxation (SDR) techniques. However, as the target and user equipment (UE) numbers increase, the computational complexity of these methods grows significantly, which can hinder real-time data processing, especially for time-sensitive LAWN applications. Beyond ISAC, Chen $et$ $al.$ \cite{chen2024LAWN} investigate the fundamental trade-off among sensing, communication, and WPT by characterizing the Cramér-Rao bound (CRB)-rate-energy (C-R-E) region, where the resulting optimization problem is reformulated in the dual domain and solved using a gradient-based method. However, the C-R-E region analysis is based on a simplified scenario with a single sensing target, a single communication UE, and a single energy-harvesting device, while more general cases involving multiple UEs and targets require further investigation.

To efficiently coordinate multiple functions in LAWNs, a promising direction is to reformulate the underlying non-convex optimization problems by exploiting the topological structure of the network. Such topology-aware (TA) solutions have been widely adopted in terrestrial networks \cite{wan2020top, yu2022topological}, satellite communication network \cite{yu2026topology}, and have demonstrated strong effectiveness in scenarios with massive connectivity \cite{wu2020massive}. Recent studies have employed TA frameworks to improve the efficiency of UAV-based networks. In \cite{hong2020formation}, a TA scheme is developed to support mobility-adaptive routing and formation control in UAV swarm networks. Cui $et$ $al.$ \cite{cui2022FANETs} suggest a UAV networking protocol that accurately captures topology changes with minimal overhead and facilitates resilient routing in UAV swarms. To further enhance TA schemes, graph neural networks (GNNs) can be employed, as the graph-based representation of nodes and edges naturally aligns with the message-passing structure of GNNs, facilitating efficient learning over complex network topologies. Adopting GNN as an observer to capture and exploit the underlying network topology, enhanced multi-UAV trajectory planning is demonstrated in \cite{du2024GNN}. Wang $et$ $al.$ \cite{wang2025GNN} consider a multi-UE multi-input single-output (MISO) UAV communication system and design a two-stage GNN-based model to maximize the energy efficiency of each served UE. In \cite{Mughal2026GNN}, Mughal $et$ $al.$ propose a generalizable TA GNN-based intrusion detection system for UAV communication systems, and its effectiveness is validated using a real-world testbed. While TA schemes have been investigated for UAV-assisted networks, the existing solutions typically lack the capability to jointly support multiple functions and coordinate heterogeneous mobile devices across terrestrial and non-terrestrial domains.

\vspace{-0.2cm}

\begin{table*}[t]
\centering
\vspace{-0.5em}
\caption{Contrasting our contributions to the state-of-the-art (SoTA) methods}~\label{table:comparison}
\vspace{-0.5em}
\small
\begin{tabular}{|m{3.8cm}|m{2cm}|m{2cm}|m{2cm}|m{2cm}|m{2cm}|}
	\hline
    
        \centering\textbf{Functionality} 
        &\centering \textbf{This work}
        &\centering \cite{MO_SC2, Wu2026LAWN}
        &\centering \cite{chen2024LAWN}
        &\centering \cite{hong2020formation, cui2022FANETs, du2024GNN, wang2025GNN}    
        &\centering \cite{Mughal2026GNN}
        \cr
        \hline         
        \hline
        
        \centering Signaling and control
        &\centering \checkmark
        &\centering \checkmark
        &\centering \checkmark
        &\centering \checkmark          
        & \centering \checkmark
        \cr 
        \hline

        \centering Target and object sensing  
        &\centering \checkmark
        &\centering \checkmark
        &\centering \checkmark
        &\centering \tikzxmark   
        & \centering \checkmark
        \cr
        \hline
        
        \centering UAV or UE charging         
        &\centering \checkmark
        &\centering \tikzxmark
        &\centering \checkmark
        &\centering \tikzxmark        
        & \centering \tikzxmark
        \cr
        \hline

        \centering TA-based coordination 
        &\centering \checkmark 
        &\centering \tikzxmark
        &\centering \tikzxmark
        & \centering \checkmark 
        & \centering \checkmark 
        \cr 
        \hline
\end{tabular}
\vspace{-1em}
\label{Contribution}
\end{table*}

\subsection{Contributions}

To satisfy the multi-functional requirements of multi-UE heterogeneous networks (HetNets), this paper proposes a TA-based framework tailored for LAWNs, which facilitates the joint optimization of sensing, communication, and charging objectives within a unified structure. In contrast to conventional approaches that solely rely on complicated multi-objective formulations and optimization procedures, as outlined in Table \ref{table:comparison}, the proposed methodology adopts a unified graph-based representation of the entire HetNet. The abstraction of the LAWN allows all functional objectives to be expressed through common graph elements, which can then be efficiently solved using basic optimization algorithms. The main contributions of this paper are summarized as follows:

\begin{itemize}
    \item \textit{TA Framework and its Graph Representation}: We devise a unified TA framework for joint terrestrial and LAWN multi-UE systems, where the network is modeled as a bipartite graph constructed based on channel strength and selectively established access point (AP)-to-UE connections. By defining node and edge attributes, heterogeneous functional objectives can be incorporated into a unified topological structure. Within this graph-based formulation, communication, sensing, and charging objectives are represented through common elements, such as profit and cost associated with nodes and edges. This abstraction explicitly characterizes the trade-off between performance gains and structural costs, demonstrating how multiple functional requirements can be jointly accommodated within a single, coherent topology.
    
    \item \textit{Low-Complexity TA-based Solution}: By leveraging the unified topological representation of the LAWN, we develop a low-complexity algorithm based on a unified resource-adjustment rule that allows topology reconfiguration through the modification of either the resource provider or the resource consumer. Specifically, the algorithm iteratively updates node activation states and associated edge relationships to adapt the network structure according to functional requirements. This design eliminates the need for objective-specific heuristic tuning, reduces computational overhead, and preserves structural consistency across heterogeneous functions in the HetNet, thereby enhancing scalability and adaptability in large-scale dynamic deployments.

    \item \textit{Comprehensive Numerical Validation}: Performance evaluations are conducted to validate the effectiveness of the proposed TA scheme and to demonstrate the advantages of modeling the LAWN as a graph. Extensive simulations under diverse network configurations and UE densities show that our TA-based method consistently outperforms SoTA approaches in terms of system efficiency and multi-functional coordination capability. In addition, the developed framework maintains robust and scalable performance in large-scale deployments involving massive terrestrial and non-terrestrial UEs, highlighting its suitability for highly dynamic and heterogeneous LAWN.
\end{itemize}
The notations used throughout this paper are summarized in Table \ref{tab:system_parameters}. The remainder of this paper is organized as follows. Section \ref{sec:SystemModel} introduces the system model of the considered LAWN. Section \ref{sec:TA-Representation} presents the TA representation of the LAWN, followed by the proposed solution in Section \ref{sec:Algorithm}. Simulation results are provided in Section \ref{sec:simulation}, and finally, conclusions are drawn in Section \ref{sec:conclusion}.

\begin{table*}[t]
    \centering
    {
    \caption{Nomenclature}~\label{tab:system_parameters}
    \vspace{-0.15cm}
    \resizebox{17.5cm}{!}
    {
    \begin{tabular}
    {p{0.8cm}|p{5.5cm}|p{0.8cm}|p{5.5cm}|p{0.8cm}|p{5.5cm}}
    \hline
    \hline
    Symbol & Meaning & Symbol & Meaning & Symbol & Meaning \\
    \hline
    $|\Xc|$ & cardinality set & $\xv^\T$ & transpose & $\xv^\herm$ & conjugate \\
    $\otimes$ & Kronecker product &
    $[\xv]_i$ & $i$-th element of the vector $\xv$ &
    $[\Xm]_{i,j}$ & $(i,j)$ element of $\Xm$  \\
    $M$ & number of APs & $M_u$ & number of UAV APs &
    $M_t$ & number of terrestrial APs \\ 
    $M_a$ & number of antennas at the AP & $K_P$ & number of charging UEs & $K_C$ & number of communication UEs \\
    $g_{mk}$ & small-scale fading between UE $k$ and AP $m$ & $\hv_{mk}$ & channel between UE $k$ and AP $m$ & $\beta_{mk}$ & path loss between UE $k$ and AP $m$ \\
    $d_x/d_y$ & inter-element spacing & $\beta_{\tau}$ & channel strength threshold for graph construction & $s_k$ & pilot signal for UE $k$ \\
    $\av(\theta,\phi)$ & steering vector  & $\rho_m$ & transmit power of AP $m$ & $N_{AP,in}$ & threshold: number of communication/sensing UEs connected to AP\\
    $\sv_{m,\text{d}}$ & downlink sending signal of $m$-th AP & $\pv_{mk_c}$ & communication precoder & $\pv_{mk_p}$ & charging precoder \\ 
    $N_\rho$ & threshold: number of UE connections to AP &
    $\rho_{m,\text{T}}$ & TA-representation: transmit power of AP $m$ & $E_{k_p,\text{T}}$ & TA-representation: received energy \\
    $m_{s}$ & selected AP for sensing target & $\tau_{s,\text{T}}$ & TA-representation: sensing threshold & $R_{\text{sum}}$ & sum rate \\
    $\tau_{p,\text{T}}$ & TA-representation: power threshold & $\theta, \phi$ & azimuth and elevation angles & $\tau_{c,\text{T}}$ & TA-representation: communication threshold \\
    $\gamma_{k}$ & SINR of UEs & $\Psim$ & cost of sensing and communication UEs & $P_{\min}$ & minimum detectable power \\
    $\Ec$ & edge set of bipartite graph & $\Am$ & adjacency matrix of graph $\Gc$ & $\Wm$ & weight matrix of graph $\Gc$ \\
    $\uv$ & $K_C\times 1 $ vector, UE nodes status & $\vv$ & $M\times 1 $ vector, AP nodes status & $N$ & maximum number of strongest UEs to each AP\\
    $\phiv_c$ & profit of communication UEs &
    $\phi_s$ & profit of sensing target & $\phiv_p$ & profit of charging UEs \\
    {$\qv_c$} & { TA-based   communication objective} & {$q_s$} & {TA-based sensing objective} & {$\qv_p$} & {TA-based charging objective} \\
    $\vv_s$ & AP selection vector of sensing target& $\vv_{p,k_p}$ & AP selection vector of $k_p$-th charging UE &$\nv_{\text{AP}}$  & $M\times 1$ vector, the number of connectivity of all APs  \\
    \hline
    
    \end{tabular}}}
\vspace{-0.1cm}
\end{table*}

\section{System Model}
\label{sec:SystemModel}

\begin{figure}
    \centering
    \includegraphics[width=\linewidth]{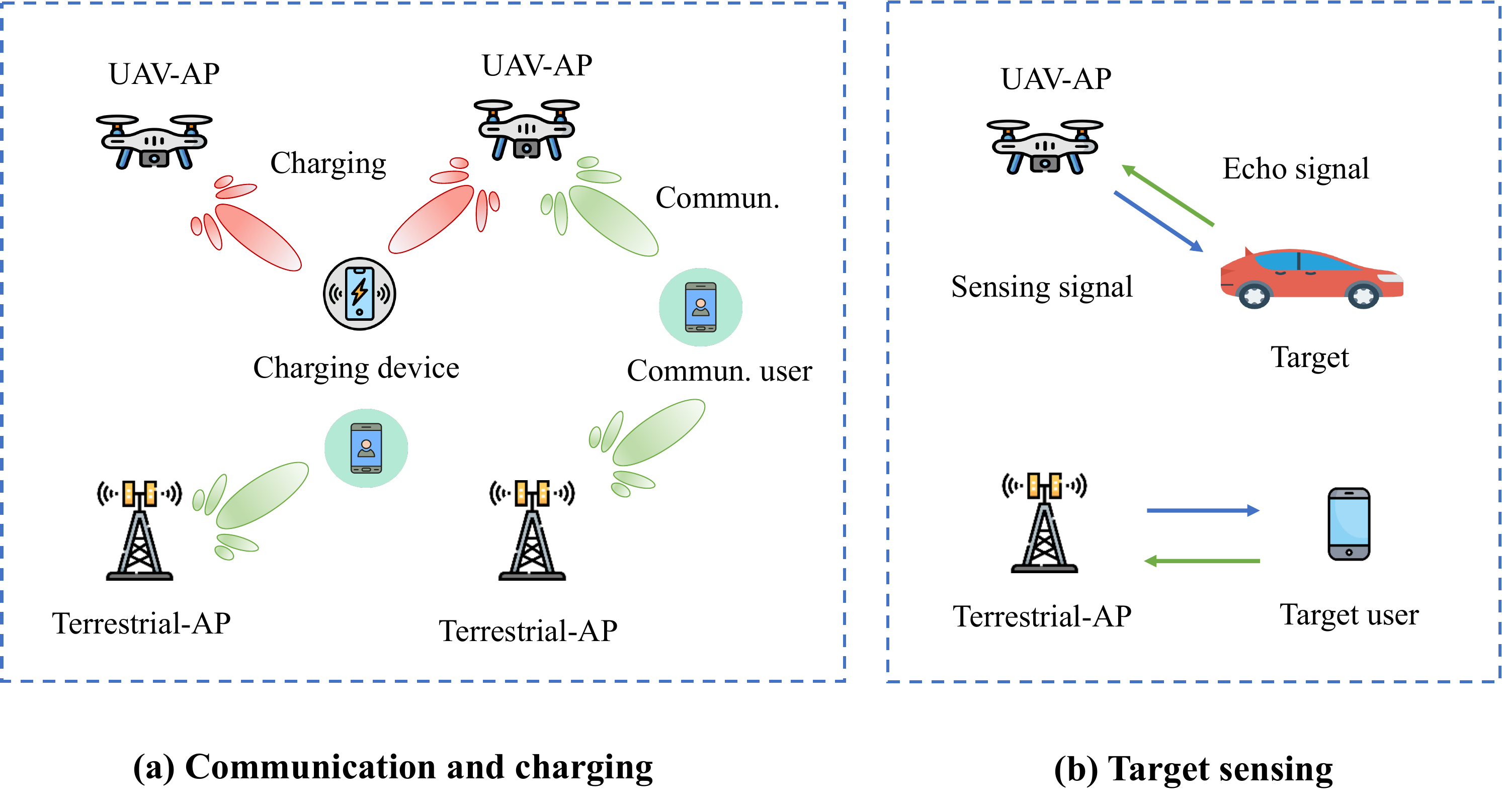}
    \caption{Multi-functional LAWN. }
    \label{fig:SystemModel}
\vspace{-0.5cm}
\end{figure}

In this paper, we consider a multi-UE LAWN consisting of both terrestrial and non-terrestrial UEs. As shown in Fig. \ref{fig:SystemModel}, $M_t$ terrestrial APs and $M_u$ UAV APs are randomly distributed within the operational area. Each terrestrial or non-terrestrial AP can perform different tasks, including communication, sensing, and WPT. Let the numbers of UEs requiring communication, sensing, and charging services from the APs be $K_C$, $K_S$, and $K_P$, respectively. The total numbers of APs and UEs are: 
\begin{equation}
    M = M_t + M_u,
\end{equation}
\begin{equation}
    K = K_C + K_S + K_P.
\end{equation}
Moreover, we denote $\Kc_C$, $\Kc_S$, and $\Kc_P$ as the communication, sensing, and charging UE sets, respectively, while $\Mc$ and $\Kc$ represent the AP and UE sets. Specifically, $|\Mc| = M$, $|\Kc| = K$, and
\begin{equation}
    \Kc = \Kc_C \cup \Kc_S \cup \Kc_P.
\end{equation}
All APs are equipped with uniform planar arrays (UPAs) with $M_a = M_{ay} \times M_{ax}$ antenna elements. We now describe the uplink phase used for channel estimation, followed by the downlink phase supporting different functional tasks in the multi-functional LAWN.

\vspace{-0.5cm}

\subsection{Uplink Phase}

To facilitate diverse functionalities in the LAWN, accurate channel estimation for all potential UEs shall be performed during the uplink phase. The received signal at AP $m$, regardless of whether it is terrestrial or UAV-based, is:
\begin{align}
    \yv_m^{\text{u}}=\sum_{k=1}^K \hv_{mk} s_k+\nv, \quad k\in\Kc,
\end{align}
where $s_k$ is the pilot of the $k$-th UE, $\nv\sim \CN(\zero,\sigma^2\IM_{M_a})$ is the additive white Gaussian noise (AWGN) vector, and $\hv_{mk}$ is the channel between the $m$-th AP and the $k$-th UE as:
\begin{align}
\hv_{mk}=\sqrt{\beta_{mk}}g_{mk}\av(\theta_{mk},\phi_{mk}),
\end{align}
where $\beta_{mk}$ and $g_{mk}$ denote the large-scale fading (i.e., path loss) and small-scale fading (i.e., $g_{mk} \sim \CN(0,1)$) between the $m$-th AP and the $k$-th UE, respectively; $\av(\theta_{mk},\phi_{mk})$ represents the UPA steering vector, which is expressed as:
\begin{align}
    &\av(\theta_{mk},\phi_{mk}) = \av_y(\theta_{mk},\phi_{mk})\otimes \av_x(\theta_{mk},\phi_{mk}), 
\end{align}
where
\begin{align}
    &\av_y(\theta,\phi) \notag \\
    & = [1,e^{\frac{2\pi d_y}{\lambda}\sin(\phi)\sin(\theta)},\dots,e^{\frac{2\pi d_y(M_{ay}-1)}{\lambda}\sin(\phi)\sin(\theta)}]^{\herm},
\end{align}
\begin{align}
    &\av_x(\theta,\phi) \notag \\
    & = [1,e^{\frac{2\pi d_x}{\lambda}\sin(\phi)\cos(\theta)},\dots,e^{\frac{2\pi d_x(M_{ax}-1)}{\lambda}\sin(\phi)\cos(\theta)}]^{\herm},
\end{align}
while $\theta_{mk}$ and $\phi_{mk}$ are the azimuth and elevation angles; $d_x/d_y$ and $\lambda$ are inter-element spacing and signal wavelength, respectively. In this study, we focus on the design of the TA-based framework and consider a one-shot channel estimation procedure, where the pilot $s_k$ is binary. This implies that certain UEs are deactivated in a given slot to mitigate interference. These UEs can be scheduled in subsequent slots using the same mechanism. Thus, this assumption does not compromise the practicality or generality of the proposed framework.

\subsection{Downlink Phase}

During the downlink phase, three categories of UEs are considered: communication, sensing, and charging UEs.\footnote{It is noted that the proposed scheme is not limited to these UE types and can be readily extended to accommodate additional categories of UEs.} {
In this paper, the APs serve multiple communication and charging UEs. Regarding sensing, we assume that multiple sensing targets are present in the operating area, while only one sensing target, noted as $k_s$, is selected for detection in each time slot.
 Thus, each AP transmits the symbol with precoder that includes communication, sensing, and charging precoders as:
\begin{align}
    \sv_{m,\text{d}}=\sum_{k_c\in\Kc_C}\pv_{mk_c}x_{k_c}+\pv_{mk_s}x_{k_s}+\sum_{k_p\in\Kc_P}\pv_{mk_p}. 
\end{align}}
Therefore, the received signal at the communication UE is:
\begin{align}
\label{eq-yc}
    y_{k_c} &=\sum_{m=1}^M\sqrt{\rho_m}\hv_{mk_c}^\T \sv_{m,\text{d}} \notag  \\
    &= \sum_{m=1}^M \sqrt{\rho_m}(\underbrace{\hv_{mk_c}^\T \pv_{mk_c}x_{k_c}}_{\text{Desired Message}} \notag \\
    &+\underbrace{\sum_{k_c^{'}\neq k_c,k_c^{'}\in\Kc_c}\hv_{mk_c}^\T\pv_{mk_c^{'}}x_{k_c^{'}}}_{\text{Communication Interference}} \notag \\ &+\underbrace{\hv_{mk_c}^\T\pv_{mk_s}x_{k_s}}_{\text{Sensing Interference}}  +\underbrace{\sum_{k_p\in\Kc_p}\hv_{mk_c}^\T \pv_{mk_p}}_{\text{Charging but No Interference}})+\underbrace{n}_{\text{Noise}},
\end{align}
where $\pv_{mk}$ is the precoder of the $m$-th AP for communication, sensing, or charging, and $k\in\Kc,\ \sum_{k}|\pv_{mk}|^2\leq 1$. {
The transmit power for the AP $m$ is denoted by $\rho_m$, which is lower than the maximum transmit power $\rho_{\max}$, while $n$ is the AWGN with variance $\sigma^2$.} Regarding the sensing UE, we have
\begin{align}
\label{eq-ys}
    y_{k_s}& = \underbrace{\sqrt{\rho_{m_s}}\hv_{m_sk_s}^\T\pv_{m_sk_s}x_{k_s}}_{\text{Desired Sensing Signal}} \notag \\
    &+\underbrace{\sum_{m=1}^M\sqrt{\rho_m}\sum_{k_p\in\Kc_P}\hv_{mk_s}\pv_{mk_p}}_{\text{Charging but No Interference}} \notag \\
    &+\underbrace{\sum_{m=1}^M \sqrt{\rho_m}\sum_{k_c\in\Kc_C}\hv_{mk_s}^\T\pv_{mk_c}x_{k_c}}_{\text{Communication Interference}}+n. 
\end{align}
Note that a single sensing target is involved, while the sensing target selects a single AP as its serving AP \cite{single-anchor}, denoted by $m_s$. In addition, the sensing task considered in this work focuses on target detection rather than target localization. Nevertheless, a serving AP is capable of determining the target position by exploiting both time-of-arrival (ToA) and angle-of-arrival (AoA) measurements. Interested readers are referred to \cite{he2025LMAC} for further details.

For a charging UE, as no information decoding is required, the entire received signal can be utilized for energy harvesting. Thus, the received signal is modeled as:
\begin{align}
\label{eq-yr}
   &y_{k_p} =n+\sum_{m=1}^M\sqrt{\rho_m}\hv_{mk_p}^\T\times \notag \\ &\left(\sum_{k_c\in\Kc_c}\pv_{mk_c}x_{k_c} +\pv_{mk_s}x_{k_s}+\sum_{k_p\in\Kc_p}\pv_{mk_p}\right). 
\end{align}

To evaluate the performance of communication, sensing, and power transfer, we consider three performance metrics: (i) the sum rate of the communication UEs, (ii) the signal-to-interference-plus-noise ratio (SINR) for the sensing task, and (iii) the received energy. For communication services, the sum rate of the communication UEs can be computed by using \eqref{eq-ys}, which is given by: 
\begin{align}
    R_{\text{sum}} &= \sum_{k_c\in\Kc_c}R_{k_c} = \sum_{k_c\in\Kc_c}\log_2(1+\gamma_{k_c}),
\end{align}
where $\gamma_{k_c}$ is the SINR of the $k_c$-th communication UE, i.e.:
\begin{align}
    \gamma_{k_c} &= \frac{\left|\sum_{m=1}^M\sqrt{\rho_m}\hv_{mk_c}^\T\pv_{mk_c}\right|^2}{\sum_{k_c^{'}\in\Kc_c,k_c^{'}\neq k_c}\text{IC}_{k_c, k_c^{'}}+\text{IS}_{k_c,s}+\sigma^2},
    \label{eq:SINR}
\end{align}
\begin{align}
    \text{IC}_{k_c,k_c^{'}}&=\left|\sum_{m=1}^M\sqrt{\rho_m}\hv_{mk_c}^\T\pv_{mk_c^{'}}\right|^2, \\
    \text{IS}_{k_c,s} &= \left|\sum_{m=1}^M\sqrt{\rho_m}\hv_{mk_c}^\T\pv_{mk_s}\right|^2. 
\end{align}
Here, $\text{IC}_{k_c,k_c^{'}}$ and $\text{IS}_{k_c,s}$ are the interferences from the $k_c^{'}$-th  communication UE and the sensing target to the $k_c$-th communication UE, respectively. 

To determine whether a target is present within a certain area, we generally evaluate the sensing SINR, defined as:
\begin{align}
    \gamma_{k_s} = \frac{\rho_{m_s}|\hv_{m_sk_s}^\T\pv_{m_sk_s}|^2}{\sum_{k_c\in\Kc_c}\text{IC}_{k_s,k_c}+\sigma^2}.
\end{align}
When the measured sensing SINR is greater than or equal to a prescribed threshold, the target is deemed detected.

For charging services, we characterize the service quality in terms of the received energy at each charging UE, which is:
\begin{align}
    \!\!\!E_{k_p}=\sum_{k_c\in\Kc_c}\text{IC}_{k_p,k_c}\!+\!\text{IS}_{k_p,s} \!+\!\left|\sum_{m=1}^M\sqrt{\rho_m}\hv_{mk_p}^\T\pv_{mk_p}\right|^2,
\end{align}
where $\text{IS}_{k_p,s}$ and $\text{IC}_{k_p,k_c}$ are defined analogously to $\text{IC}_{k_c,k_c^{'}}$ and $ \text{IS}_{k_c,s}$ in \eqref{eq:SINR}. Based on this configuration, we next present a graph-theoretic representation of the LAWN as follows.

\section{Topology Aware Representation}
\label{sec:TA-Representation}

\begin{figure*}
    \centering
    \includegraphics[width=0.8\linewidth]{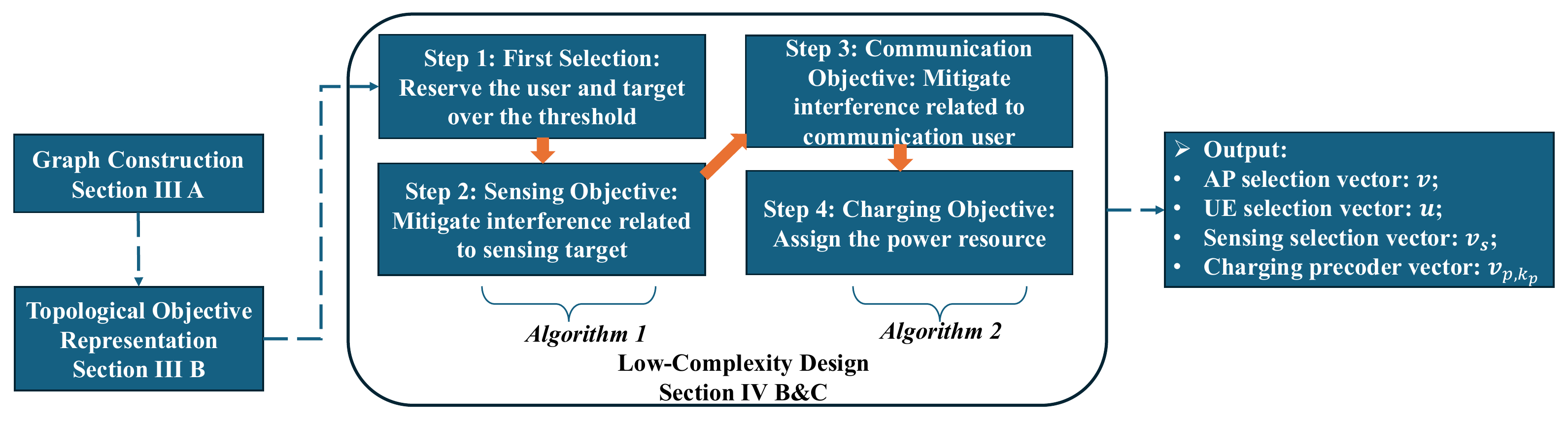}
    \caption{Workflow of the proposed TA-based low-complexity framework starting from graph construction and ending with the output of the resource and selection strategy.}
    \label{fig:flowgraph}
    \vspace{-0.4cm}
\end{figure*}

Figure \ref{fig:flowgraph} illustrates that there are three main components in the proposed framework: graph-based reconstruction, objective representation, and the TA-based algorithm design. This section focuses on the first two components, namely, the construction of the bipartite graph for LAWN modeling and embedding of multiple objectives into the graph.

\vspace{-0.2cm}

\subsection{Graph Representation of LAWN}

\begin{figure}
    \centering
    \includegraphics[width=0.8\linewidth]{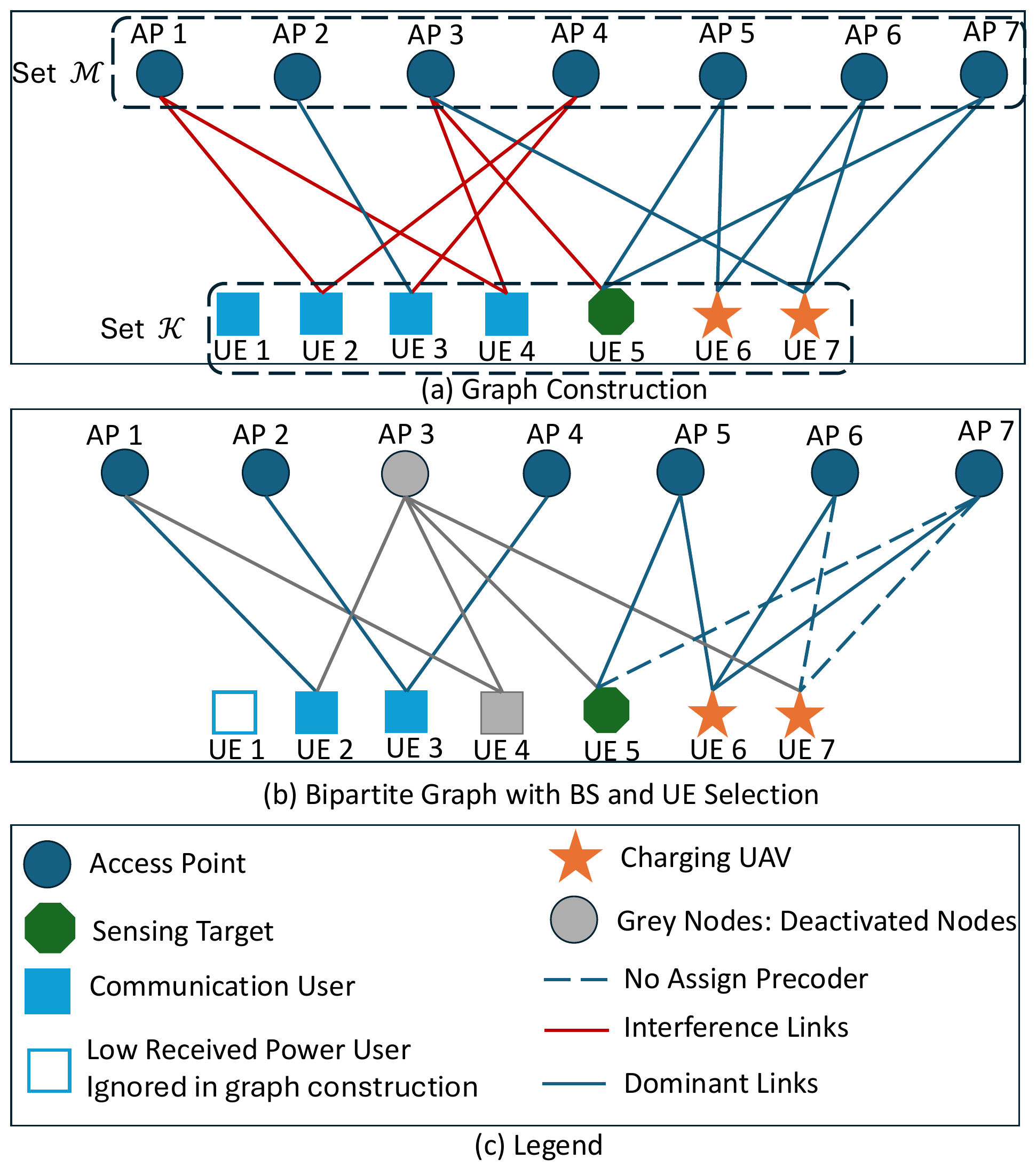}
    \caption{Graph construction and graph-based resource adjustment: (a) bipartite graph construction of the multi-functional heterogeneous LAWN based on channel strengths and feasible AP-UE associations; (b) graph reconfiguration after applying the resource-adjustment strategy, where selected nodes or links are deactivated to satisfy communication, sensing, and charging constraints; (c) legend for the node and link types.}
    \label{fig:BipartiteGraph}
    \vspace{-0.5cm}
\end{figure}

In the heterogeneous LAWN, UEs performing specific functions, such as sensing, communication, or power transfer, are associated with either terrestrial or non-terrestrial APs. The fast signal attenuation and blockage effects may lead to the links between the serving APs and the UEs being too weak to be considered, thereby resulting in a sparse network. Therefore, the LAWN can be represented by a bipartite graph $\Gc=(\Mc,\Kc,\Ec)$. As depicted in Fig. \ref{fig:BipartiteGraph}, the solid lines represent the connections between APs and UEs, which are established according to the following two rules:
\begin{itemize}
    \item \textit{Rule 1: \textbf{Channel Strength-based Selection}} \\
    The channel strength is commonly used to determine whether a link between an AP and a UE is feasible in communication systems, and the corresponding channel strength threshold is defined as: 
    \begin{align}
        a_{mk} = \begin{cases}
            0, &  \beta_{mk} < \beta_{\tau}, k\in\Kc, \\
            1, &  \beta_{mk}\geq \beta_{\tau},k\in\Kc,
        \end{cases}
    \end{align}
    where $a_{mk}$ denotes the $(m,k)$ column element of the adjacency matrix $\Am$ of $\Gc$. It is important to note that the channel strength threshold $\beta_{\tau}$ significantly affects performance. This threshold can be controlled by comparing it with the noise strength, and further details will be discussed in Section \ref{sec:simulation}. 
    
    \item \textit{Rule 2: \textbf{Dominant Connectivity Selection}} \\
    Considering the limited capacity and computational resources of the serving AP, each AP selects $N$ strongest channels as the key connections. The corresponding entries in $\Am$ that also satisfy Rule 1 are set to `1', indicating that the associated links are retained in the bipartite graph.
\end{itemize}

Based on Rules 1 and 2, the bipartite graph for the LAWN is constructed. In the following, we provide further insights into the resulting bipartite graph:

\begin{itemize}
    \item Figure \ref{fig:BipartiteGraph}(a) illustrates the bipartite graph representing the connectivity between the APs and the UEs. In this graph, $\Mc$ denotes the set of AP nodes, $\Kc$ denotes the set of UE nodes, while $\Ec$ denotes the set of connectivity edges between APs and UEs.
    
    \item
    We define the corresponding weight matrix as $\Wm$. {
    Its $j$-th column vector is denoted as $\wv_j$, while the $i$-th row and $j$-th column element is defined as $w_{ij}$.} An edge between AP $m$ and UE $k$ is established if the corresponding large-scale channel gain exceeds a predefined threshold. Accordingly, the weight of each edge is defined as $w_{mk}=a_{mk}\beta_{mk}$.
    
	\item For both communication and sensing UEs, activating multiple links associated with the same AP simultaneously may cause strong inter-UE interference. Our objective is to preserve the dominant links while mitigating the severe interference induced by conflicting connections among dominant connections.
    
    \item During the uplink channel estimation phase, the pilot signals associated with charging UEs may introduce interference to both communication and sensing UEs, since multiple users may share the same AP resources and cause pilot contamination. {
    However, since the number of charging UEs is much smaller than that of communication UEs, the impact of charging UEs pilot contamination on channel estimation is expected to be relatively limited. To verify this assumption, Section \ref{sec:simulation} compares the communication performance with and without pilot contamination caused by the charging UEs.} In contrast, during the downlink transmission phase, the charging waveform is generated and controlled by the network and its associated channels are estimated for precoder design. As a result, the charging-induced signal component becomes predictable and can be mitigated (e.g., through cancellation based on the known waveform and estimated equivalent channel). Therefore, charging-UE links are not treated as stochastic interference sources to communication or sensing UEs in the downlink phase, nor are they significantly affected by communication or sensing transmissions under the considered structural regulation.

\item In this work, we distinguish between structural interference and estimation-induced interference. Structural interference refers to the direct signal-level interference caused by simultaneous transmissions under a given AP-UE activation structure, which is determined by large-scale channel characteristics and selection decisions.
Although pilot contamination and channel estimation errors are also influenced by user associations at the APs, they affect system performance indirectly through imperfect precoding rather than direct signal superposition. Therefore, they belong to estimation-induced interference rather than structural interference.
Since the number of charging UEs is relatively small in the considered scenario, the estimation-induced interference from charging users is negligible. As a result, we focus on structural interference management in this framework.
\end{itemize}

Once the graph representation of the LAWN is established, we next describe how different functional objectives can be jointly represented using the bipartite graph.

\vspace{-0.5cm}

\subsection{Topological Objective Representation}

In constructing the topological representation, a simplified maximum ratio transmission (MRT)-based precoding model is adopted without normalization, and the noise power term $\sigma^2$.
By leveraging the average power allocation strategy, the communication objective in the topological structure can be approximated by:
\begin{align}
    \gamma_{k_c,\text{T}} &=  \frac{\sum_{m=1}^Mv_m\rho_{m,\text{T}}a_{mk_c}\beta_{mk_c}^2}{\sum_{\substack{k_c^{'}\in\Kc_C \\  k_c^{'}\neq k_c}} \text{IC}_{k_c,k_c^{'},\text{T}}+\rho_{m_s,\text{T}}a_{m_sk_c}\beta_{m_sk_c}a_{m_sk_s}\beta_{m_sk_s}} \notag \\
    &=\frac{\sum_{m=1}^Mv_m\rho_{m,\text{T}}w_{mk_c}^2}{\sum_{k_c^{'}\in\Kc_C,k_c^{'}\neq k_c}\text{IC}_{k_c,k_c^{'},\text{T}}+\rho_{m_s,\text{T}}w_{m_sk_c}w_{m_sk_s}},
\end{align}
where
\begin{align}
    &\text{IC}_{k_c,k_c^{'},\text{T}} = u_{k_c^{'}}\sum_{m=1}^Mv_m\rho_{m,\text{T}}a_{mk_c}\beta_{mk_c}a_{mk_c^{'}}\beta_{mk_c^{'}},\notag\\
    &\qquad\qquad =u_{k_c^{'}}\sum_{m=1}^Mv_m\rho_{m,\text{T}}w_{mk_c}w_{mk_c^{'}}, 
\end{align}
while the sensing and charging objectives are:
\begin{align}
    &\gamma_{k_s,\text{T}}= \frac{\rho_{m_s,\text{T}}\beta_{m_sk_s}^2}{\sum_{k_c\in\Kc_C}\text{IC}_{k_c,k_s,\text{T}}}=\frac{\rho_{m_s,\text{T}}w_{m_sk_s}^2}{\sum_{k_c\in\Kc_C}\text{IC}_{k_c,k_s,\text{T}}},\\
    &E_{k_p,\text{T}} = \sum_{k_c\in\Kc_C}\text{IC}_{k_c,k_p,\text{T}}+\rho_{m_s,\text{T}}a_{mk_p}\beta_{mk_p}a_{mk_s}\beta_{mk_s} \notag\\
    &\qquad +\sum_{m=1}^Mv_{p,mk_p}\rho_{m,\text{T}}a_{mk_p}\beta_{mk_p}^2\notag\\
    &\qquad = \sum_{k_c\in\Kc_C}\text{IC}_{k_c,k_p,\text{T}}+\rho_{m_s,\text{T}}w_{m_sk_p}w_{m_sk_s} \notag\\
    &\qquad +\sum_{m=1}^Mv_{p,mk_p}\rho_{m,\text{T}}w_{mk_p}^2.
\end{align}
Here, $\rho_{m,\text{T}}$ denotes the allocated power based on TA in the graph; $\text{IC}_{k_c,k_c^{'}, \text{T}},\text{IC}_{k_c,k_s,\text{T}}, \ \text{and } \ \text{IC}_{k_c,k_p,\text{T}}$ are the corresponding interference components in the TA-based representation with four sets of binary decision variables to be controlled.  Specifically, the binary vector $\uv\in\CC^{K_C\times 1}$ indicates whether the UE is served in the current slot, and $\vv\in\CC^{M\times 1}$ indicates the AP status. The variable $\vv_s$ denotes the AP selection vector for the sensing target, where $v_{s,m}$ is the $m$th element, subject to $\sum_{m=1}^M v_{s,m} = 1$ and $v_{s,m} \leq v_m$, and $m_s$ denotes the AP selected for the sensing target. Moreover, $\vv_{p,k_p}$ denotes the precoder assigned vector to the $k_p$-th charging user, and $v_{p,mk_p}$ is the $m$-th element of the vector, and we have  $v_{p,mk_p}\leq v_m$. We note that the sensing task is mandatory in our model, i.e., the sensing target is always active. By contrast, a charging UE may not be assigned any dedicated charging precoder. This captures the practical case in which UEs with severely attenuated channels cannot satisfy the minimum received-power requirement for effective charging, e.g., $E_{k_p} < P_{\min}$, and are therefore not treated as primary charging UEs, although they may still harvest negligible leakage energy.

\section{Low-Complexity Algorithm Design}
\label{sec:Algorithm}

Through the proposed topology-based representation, the communication, sensing, and charging objectives are no longer treated as isolated metrics. Instead, they are unified within a common structural framework, where each objective is determined by the interplay among the desired signal, interference, and allocated power. In this section, we demonstrate how these three objectives can be uniformly characterized by three fundamental components. This unified formulation further reveals that the overall system performance is essentially governed by the aforementioned binary control variables, i.e., $\{u_k,v_m,v_{s,m},v_{p,mk_p}\}$. Building on this structural insight, we develop a low-complexity and analytically tractable algorithm for the multi-UE multi-functional algorithm design.

\vspace{-0.5cm}

\subsection{Profit and Cost Design}

The bipartite graph is characterized by the adjacency matrix $\Am$ and the weight matrix $\Wm$, where $\Am$ indicates the dominant connections in the graph and $\Wm$ specifies the weights assigned to the edges. The procedure for constructing the bipartite graph will be detailed in the algorithm. To exploit the structural commonality provided by the bipartite-graph formulation, we introduce two concepts, namely, \emph{Profit} and \emph{Cost}, that are shared across different objectives. Although their specific forms vary with the related objective function, they follow the same underlying principles:
\begin{itemize}
    \item \textit{\textbf{Profit}}: The desired received signal power for all types of UEs is defined as: 
    \begin{align}
        &[\phiv_c]_{k_c} = u_{k_c}\sum_{m=1}^Mv_m\rho_{m,\text{T}}w_{mk}^2, \quad k_c\in\Kc_C, \label{eq:Profit_Commun}\\
        &\phi_s = \sum_{m=1}^M\rho_{m_s,\text{T}} v_{s,m}w_{m_sk_s}^2,\label{eq:Profit_Sen} \\
        &[\phiv_p]_{k_p} = \sum_{m=1}v_m\rho_{m,\text{T}}w_{mk_p} \times \notag \label{eq:Profit_Char}\\
        & \left (\sum_{k\in\Kc_C}u_{k}w_{mk}+v_{s,m}w_{mk_s}+v_{p,mk_p}w_{mk_p} \right ).
        \end{align}
        
        \item \textit{\textbf{Cost}}: The interferences experienced by the communication and sensing targets are modeled as the cost terms: 
        \begin{align}
            &[\Psim]_{m, k_c} = v_m\rho_{m,\text{T}}\times\notag\\
            &\left(\sum_{k_c^{'}\neq k_c}u_{k_c^{'}}w_{mk_c}w_{mk_c^{'}}+v_{s,m}w_{mk_c}w_{mk_s}\right),\\ 
            &[\Psim]_{m, k_s} = v_m\rho_{m,\text{T}}\sum_{k_c\in\Kc_C}u_{k_c}w_{mk_c}w_{mk_s},
          \label{eq:Cost} 
        \end{align}
        where $\rho_{m,\text{T}}$ denotes the approximate transmit power allocated to the $m$-th AP. In this paper, an average power allocation scheme is adopted; therefore, $\rho_{m,\text{T}}$ can be expressed as
        \begin{align}
            &\rho_{m,\text{T}} =\frac{1}{[{\nv_{\text{AP}}]_m}} \label{eq:rhoT}, \\
            &[\nv_{\text{AP}}]_m = \sum_{k_c\in\Kc_C}u_{k_c}v_ma_{mk_c}+a_{mk_s}v_{s,m} \notag \\
            &\qquad\quad +\sum_{k_p\in\Kc_P}a_{mk_p}v_{p,mk_p},\label{eq:nAP}
        \end{align}
\end{itemize}

Based on the definitions of \emph{Profit} and \emph{Cost}, the communication, sensing, and charging performance metrics are characterized as:
\begin{align}
    & \text{Communication}: [\qv_c]_{k_c} = \frac{[\phiv_{c}]_{k_c}}{\sum_{m=1}^M[\Psim]_{m,k_c}}, \label{eq:FProfit_Commun}\\
    & \text{Sensing}: q_{s} = \frac{\phi_s}{\sum_{m=1}^M[\Psim]_{m,k_s}+\phi_s}, \label{eq:FProfit_Sen}\\
    & \text{Charging}: [\qv_p]_{k_p} = [\phiv_p]_{k_p}. \label{eq:FProfit_Char}\\
    &\text{For All: } q=\sum_{k_c\in\Kc_C}[\qv_c]_{k_c}+q_s+\sum_{k_p\in\Kc_P}[\qv_p]_{k_p}\label{eq:all}
\end{align}

It is observed in \eqref{eq:FProfit_Commun}, we adopt the sum rate as the objective, where the SINR plays a crucial role in quality of service (QoS) evaluation. Under the proposed \emph{Profit-Cost} framework, the desired signal component (i.e., the profit) is represented by $[\phiv_c]_{k_c}$, while the corresponding interference is quantified by the \emph{Cost} term. Accordingly, the communication objective naturally takes an SINR-based form. Similarly, the sensing SINR is adopted as the sensing performance metric, as shown in \eqref{eq:FProfit_Sen}. Since only a single sensing target is considered in this paper, the interference originates exclusively from communication UEs. Consequently, the sensing objective can also be formulated within the same \emph{Profit-Cost} framework.

For the charging objective, performance is evaluated based on the total received energy at each charging UE. {
In \eqref{eq:Profit_Char}, the charging profit accounts only for the gain contributed by the precoder intended for the $k_p$-th charging UE. Although channel correlation may allow the precoders of other charging UEs to contribute additional received power, it is designed only as a low-complexity approximation of the received charging power and is not intended to exactly represent the actual charging performance. Moreover, such correlation does not introduce a negative contribution, and ignoring it reduces the computational complexity.
This approximation is also unlikely to significantly affect the selection strategy. If two charging UEs have strongly correlated and sufficiently strong channels, both can contribute to the charging power. If one charging UE has a weak channel and fails to satisfy the charging-availability threshold, more power is allocated to the other one. When the correlation is weak, explicitly accounting for it has only a limited effect on the selection result.} Regarding the formulation, in contrast to communication and sensing tasks, interference does not degrade charging performance. Instead, all received power, including that from precoders intended for other UEs, contributes positively to charging. Hence, the charging objective depends only on the accumulated received power. Moreover, a charging UE is considered successfully served only if its received energy exceeds a predefined threshold $\tau_{p,\text{T}}$.

\subsection{Low-Complexity Algorithm I}

The subsequent subsections present the details of our solution.
As shown in the workflow graph, Fig. \ref{fig:flowgraph}, the algorithm is divided into four steps, detailed in Algorithms \ref{Alg1-1} and \ref{Alg1-2}, respectively. In this subsection, we detail the first two steps of the overall framework, which operate on the constructed graph and progressively refine the selection decisions. The following points are used to explain Algorithm \ref{Alg1-1}: 

\begin{algorithm}[t]
\small
\caption{Low-Complexity TA Framework I}
\label{Alg1-1}
\KwIn{Matrix $\Am, \Wm$, thresholds $\tau_{c,\text{T}}, \tau_{s,\text{T}}$}
\KwOut{UE, AP status $\uv$, $\vv$, sensing selection vector $\vv_s$}
Initialization: $\uv=\one_{K_C},\vv=\one_{M},\vv_s=\zero_{M}, \vv_{p,k_p} = \one_{M}$
\SetKwBlock{StepOne}{\textbf{Step 1: First selection}}{}
\SetKwBlock{StepTwo}{\textbf{Step 2: Sensing Objective}}{}

\StepOne{
\If{$\wv_{k_s}=\zero_M$}{Sensing target cannot be detected, }
Check the profit of each communication UE\;
\If{$[\phiv_c]_k < \tau_{c,\text{T}}$}{$u_k = 0$ \;}
}
\StepTwo{
Select the index corresponding to the maximum element in vector $\wv_{k_s}$ as $i=\arg\max_{i\in\Mc}[\wv_{k_s}]_i$\;
Assign: $v_{s,i}=1$, and define $m_s=i$\;
Calculate the profit, cost, and $q_s$ according to \eqref{eq:Profit_Commun}-\eqref{eq:FProfit_Char}\;
\While{$q_s < \tau_{s,\text{T}}$}{Select the worst AP according to the cost as: $i_s = \arg\max_{i_s\in\Mc}[\Psim]_{i_s,k_s}$\;
Calculate $q^{'}$ if closing the $i_s$-th AP\;
Select the communication UEs that conflict at the $i_s$-th AP as the UE set $\Kc_{C,2}$\;
Calculate $q^{''}$ if closing the UEs\;
\If{$q^{'} > q{''}$}{Update the AP status vector as $v_{i_s}=0$ and $q_s$}
\Else{Update the communication UE status vector as $u_{k} = 0,k\in \Kc_{C,2}$, $\rho_{m,\text{T}}$, and $q_s$}
}}
\Return{$\uv,\vv,\vv_{s}$}\;
\end{algorithm}

\begin{itemize}
\item Step 1 is designed to identify UEs that are feasible for communication or sensing in the current time slot. For the sensing target, it may experience extremely weak channels from all APs. In such a case, the corresponding links are naturally discarded during the graph construction stage (e.g., their weights fall below a predefined threshold and are treated as negligible), rendering the sensing target effectively isolated in the reconstructed topology. Consequently, it cannot be supported in the current topological structure.

\item For communication UEs, the situation is slightly different. Although some of their individual channel strengths may exceed the graph-construction threshold $\beta_\tau$, the overall achievable received signal strength (e.g., aggregated power from candidate APs) may still be insufficient to justify scheduling compared with other UEs. Therefore, such UEs are not selected in this slot. However, unlike the sensing case where links are structurally removed in the topology, these communication UEs remain eligible and may be reconsidered in subsequent time slots under the same topology structure.
\item Step 2 focuses on satisfying the sensing objective. In this work, only one AP is selected to transmit the sensing signal toward the target. While this single-AP design simplifies the sensing structure, it also makes the sensing SINR highly sensitive to interference from concurrently active APs and communication UEs.
In this step, we first identify the dominant sensing link, namely, the AP that provides the strongest channel connectivity to the target. Based on this selection, the represented sensing SINR, i.e., $q_s$, is evaluated under the current activation pattern. If the SINR requirement is not satisfied, the APs or communication UEs that contribute most to the interference degradation are identified. The activation status of these interfering nodes is then adjusted, and the corresponding sensing metric $q_s$ is re-evaluated.
By comparing the resulting $q_s$ values under different adjustment strategies, the configuration yielding better performance is selected. This iterative procedure continues until the SINR requirement is satisfied.
\end{itemize}

\subsection{Low-Complexity Algorithm II}

After steps 1 and 2, where infeasible links are removed and sensing requirements are satisfied, we proceed to the remaining stages of the proposed framework. The following steps focus on refining the communication strategy and performing charging resource allocation, as presented in Algorithm \ref{Alg1-2}.

\begin{algorithm}[t]
\small
\caption{Low-Complexity TA Framework II}
\label{Alg1-2}
\KwIn{Matrix $\Am, \Wm, \uv,\vv, \vv_s$, thresholds: $\tau_{p,\text{T}},N_{AP,in},N_\rho$}
\KwOut{UE status $\uv$, AP status $\vv$, sensing selection vector $\vv_s$, charging AP selection vector $\vv_{p,k_p}$}
\textbf{Initialization}: $\vv_{p,k_p} = \one_{M}$
\SetKwBlock{StepThree}{\textbf{Step 3: Communication Objective}}{}
\SetKwBlock{StepFour}{\textbf{Step 4: Charging Objective}}{}

\StepThree{
Check the connectivity of communication UE at each AP as $[\nv_{\text{AP}}^{'}]_m=\sum_{k\in\Kc_C}v_m a_{mk} u_k$\;
\While{$\max_{m\in\Mc} [\nv_{\text{AP}}^{'}]_m> N_{\text{AP},in}$}{
Select the AP as $i_m=\arg\max_{i_m} [\nv_{\text{AP}}^{'}]_{i_m}$\;
Calculate $q^{'}$ if close the $i_m$-th AP as \eqref{eq:all}\;
Select the conflict UE as $k_c = \arg\min_{k_c}[\phiv_c]_{k_c}$\;
Calculate $q^{''}$ if closing the UEs as \eqref{eq:all}\;
\If{$q^{'} > q^{''}$}{Update the AP status vector as $v_{i_m}=0$}
\Else{Update the communication UE status vector as $u_{k_c} = 0$}
Update the connectivity $\nv_{\text{AP}}^{'}$, $\rho_{m,\text{T}}$, profit, cost, and $q$.
}
}
\StepFour{
Check the charging possibility\;
\If{$[\phiv_p]_{k_p}<\tau_{p,\text{T}}$}{$\vv_{p,k_p}=\zero_{M}$}
Check the connectivity $\nv_{\text{AP}}$ as \eqref{eq:nAP}\;
\While{$\max_m [\nv_{\text{AP}}]_m > N_\rho, \forall m\in\Mc$}{
Select the AP as $i_{m}=\arg\max_{i_{m}}n_{\text{AP},i_{m}}$\;
Select the conflicting charging UE as $i_p = \arg\min_{i_p}[\phiv_p]_{i_p}$\;
Calculate $q^{'}$ if close the $i_m$-th connectivity, i.e., $v_{p,mk_p}=0$ as \eqref{eq:all} \;
Calculate $q^{''}$ if closing all the connectivity of the charging UE, i.e., $\vv_{p,i_p}=\zero_M$ as \eqref{eq:all}\;
\If{$q^{'} > q^{''}$}{Update the vector as $v_{p,i_mi_p}=0$}
\Else{Update the vector as $\vv_{p,i_p} = \zero_M$}
Update the connectivity $\nv_{\text{AP}}$, $\rho_{m,\text{T}}$, profit and cost.
}
Repeat steps from 20 to 28, and check the charging possibility as steps 16-17\;}
\Return{$\uv,\vv,\vv_{s},\vv_{p,k_p}$}\;
\end{algorithm}
The following points are used to explain the algorithm: 
\begin{itemize}
\item {
Algorithms~\ref{Alg1-1} and~\ref{Alg1-2} mainly involve two types of thresholds: channel-strength thresholds, including $\tau_{c,\text{T}}$ and $\beta_\tau$, and resource-allocation thresholds, including $\tau_{s,\text{T}}$, $N_{AP,in}$, and $N_\rho$. Specifically, $\beta_\tau$ is used to filter out weak communication UEs, while $\tau_{c,\text{T}}$ further removes communication UEs that become relatively weak during the AP-selection process. The integer threshold $\tau_{s,\text{T}}$ controls the interference at the sensing target by limiting the number of links connected to it in the constructed bipartite graph. In this paper, we set $\tau_{s,\text{T}}=1$, allowing only one such connection and thereby minimizing the interference at the sensing target.
Moreover, $N_{AP,in}$ is used to mitigate interference among communication UEs by limiting the maximum number of connections associated with each AP. We generally set $N_{AP,in}=1$, such that overlap among users is preferably avoided. Finally, $N_\rho$ determines the maximum number of users connected to each AP. A larger $N_\rho$ implies that a lower transmit power is allocated to each connected user.
}
    \item The core concept of the TA-based framework is a unified resource-adjustment rule. If the results of the current pattern fail to meet the predetermined threshold, two alternative actions may be considered: adjusting the resource provider or altering the resource consumer. The final profit for each action is calculated, and the action yielding the highest profit is chosen. This selection process is repeated until the requirements established by the thresholds in the representative TA-based framework are met. This mechanism facilitates the coordinated optimization of multiple objectives via a unified decision-making framework.
    
    \item Specifically, regarding the communication UEs and sensing targets, the two actions involve deactivating the communication UEs (resource consumers) that cause interference or the corresponding AP (resource provider). The interference between communication and sensing objectives can be mitigated to an acceptable level. For charging UEs, there is no interference, however, the precoder selected for the charging UE may influence power allocation. Consequently, the two actions involve either deactivating the specific precoder (resource provider) or the entire precoding vector (resource consumer). In essence, interference mitigation and power reallocation are both represented in a common resource reconfiguration problem, facilitating a unified and scalable design.  
\end{itemize}
{
\begin{remark}
    The computational complexity of Algorithms \ref{Alg1-1} and \ref{Alg1-2} can be reduced to a lower level through the incrementally calculation. In the Step 1, which corresponds to initial communication and sensing user checking, it requires only $\mathcal O(M+K_C)$ operations; In Steps 2 and 3, the objective values associated with deactivating an AP or communication UE can be incrementally evaluated. Each iteration requires at most $\mathcal O(MK)$ operations, and the number of iterations is upper bounded by $M+K_C$, since each iteration deactivates at least one AP or communication UE. Therefore, the complexities of Steps 2 and 3 are both $\mathcal O\left(MK(M+K_C)\right)$. In Step 4, each iteration also requires at most $\mathcal O(MK)$ operations. As each AP processes at most $N_\rho-N_{AP,in}$ charging connections, where $N_\rho$ and $N_{AP,in}$ are small fixed integers, the number of iterations is $\mathcal O(M)$, resulting in $\mathcal O(M^2K)$ complexity. Combining all steps, the overall complexity is $\mathcal O\left(MK(M+K_C)\right)$, which is upper bounded by $\mathcal O\left(MK(M+K)\right)$.
\end{remark}}
To conclude, the TA-based framework can embed multiple objectives into a unified structural representation. This framework focuses on mitigating interference and allocating power resources in a coordinated manner. Through the four sequential steps, the method enables decision-making from both the UE and AP perspectives, rather than relying solely on the UE side as in traditional approaches. Although this method includes a set of thresholds that are closely related to system performance, they can be selected in a comprehensive manner, since each threshold has a clear physical interpretation with respect to its corresponding objective. Note that our framework is not limited to this specific multi-objective problem; instead, it provides a general structural paradigm for multi-objective network optimization, offering the objectives can be formulated within a unified representation. We next evaluate the performance in terms of communication, sensing, and charging by comparing our solution with benchmark schemes.

\section{Simulation Results}
\label{sec:simulation}

\subsection{Simulation Setups}

We first describe the simulation setup and introduce the benchmark schemes used for comparison. We consider a $2\,\text{km}\times 2\,\text{km}$ operating region, where $64$ terrestrial APs are randomly deployed. In addition, $64$ UAV APs are randomly generated, with altitudes uniformly distributed between $10$ m and $50$ m. All APs are equipped with $2\times 2$ UPA antennas. A large number of UEs are randomly distributed over the region, among which $80\%$ are UAV UEs and the remaining $20\%$ are terrestrial UEs. Among the participating UAV UEs, $9$ are designated as charging UEs, and the remaining one are selected as the sensing target. To enhance charging efficiency, at least one UAV AP is located within a distance of $10$ to $50$ m from each charging UAV UE. Unless otherwise specified, the simulation parameters are provided in Table \ref{tab:simulation_parameters}. {
Additionally, to ensure a fair performance comparison, we generate 100 independent spatial deployment realizations, each consisting of randomly located APs and UEs. For each deployment realization, 100 independent small-scale fading realizations are generated. The spatial deployment determines the corresponding path loss  coefficients, based on which the optimization is carried out for all considered methods. For each realization, all schemes use the same numbers, locations, and channel realizations of the APs and different types of UEs, i.e., communication, sensing, and charging, such that the only difference lies in the adopted selection strategy.}

To evaluate the proposed scheme, we compare it with two typical benchmark schemes, namely, \emph{No Selection} and \emph{UE-Centric Selection}. In the LAWN, numerous AP to UE links exhibit weak channel conditions; accounting for these links would incur unnecessary channel-estimation overhead and computational complexity. To ensure a fair comparison, we apply a UE-centric pruning step to both benchmarks to exclude insignificant links. Specifically, each UE retains only the three APs with the largest pathloss-based channel strength, while the remaining APs are ignored. Consequently, channel estimation is not performed for the discarded links, and they are not considered for downlink transmission.

For the benchmarks, \emph{No Selection} activates all communication UEs,
while the \emph{UE-Centric Selection} scheme follows the cell-free massive multiple-input multiple-output (CF-mMIMO) principle in \cite{liu2020graph,he2025RSS}, where UE activation is determined by dominant channel relationships. 
For the sensing objective, the AP with the strongest channel to the sensing target is selected as the serving AP. For the other dominant APs associated with the sensing target, all communication UEs having dominant links to these APs are deactivated, thereby ensuring optimal sensing performance under the UE-centric structure. For communication, the selection starts from the communication UE with 3 stronger AP based on CKM \cite{CKM2026}. An UE is activated only if it does not share any dominant AP with the already selected UEs; otherwise, it is deactivated. The strongest UE is selected first, after which the next strongest UE among the remaining candidates is examined using the same criterion. This process continues until all activated communication UEs are mutually non-overlapping under the UE-centric structure. 

{
The number of served UEs varies among the considered methods and is also affected by the size of the deployment area. Under the \textit{No Selection} scheme, all UEs are activated. Consequently, many weak users consume power resources without effectively improving their own performance, which degrades the overall system performance. In the \textit{UE-Centric Selection} scheme, the selection process also depends on channel strength, allowing weak users to be filtered out. Similarly, the proposed \textit{TA-based UE and AP selection} explicitly addresses the weak-user issue by introducing channel-strength thresholds $\beta_\tau$.
The number of weak users depends on both the deployment area and the number of APs. For example, when 200 users and 128 APs are uniformly distributed over a 2 km $\times$ 2 km area, only approximately 50 users satisfy the threshold. When the area is reduced to 1 km $\times$ 1 km, approximately 120 users satisfy the threshold. Among the remaining communication users, only a subset is selected in each time slot to mitigate mutual interference. Users that are not served in the current one-shot time slot can be scheduled in subsequent time slots. Note that since our method jointly performs UE and AP selection and allows interfering APs to be deactivated, it can generally serve more UEs than the \textit{UE-Centric Selection} scheme.
}

\begin{table}[t]
\centering
\caption{Simulation Parameters}
\label{tab:simulation_parameters}
\begin{tabular}{c|c|c|c}
\hline
\textbf{Parameter} & \textbf{Value}&\textbf{Parameter} & \textbf{Value} \\
\hline
\hline
$M_u$ & 64 & $M_t$ & 64 \\\hline
$K_P$ & 9 & $K_S$ & 1\\\hline
$f$ & 2.6 GHz &$W$ & 100 MHz \\\hline
 $P_{\min}$ & -20 dBm &$\rho_{\max}$  & 5 W\\\hline
$M_a$ & $2\times 2$&$N$ & 3\\\hline
$\sigma^2$ & -90 dBm& $\tau_{p,\text{T}}$ &$5\times10^{-6}$\\\hline
$\tau_{c,\text{T}}$& $10^{-11}$&$\tau_{s,\text{T}}$ & 1\\\hline
$N_{AP,in}$ &1&$N_{\rho}$ &3\\
\hline
\end{tabular}
\end{table}

\vspace{-0.4cm}

\subsection{Communication Performance}

\begin{figure}
    \centering   
    \includegraphics[width=0.75\linewidth]{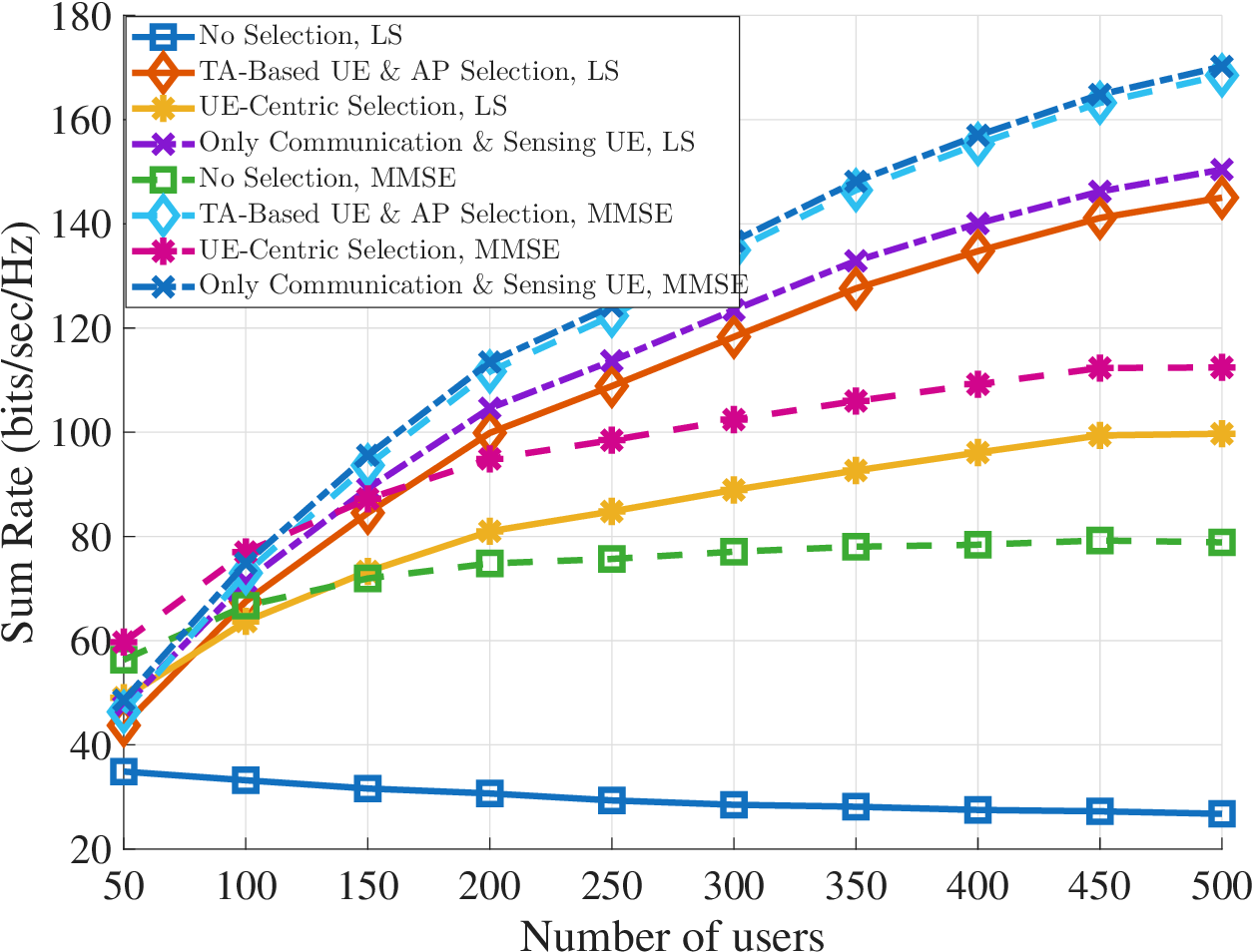}
    \caption{Sum rate versus massive number of UEs}
    \label{fig:sumrate}
\end{figure}

We start by evaluating the sum-rate performance of the proposed TA-based scheme. Channel estimation is performed only for the dominant links between the activated APs and UEs. We consider both least squares (LS) and local minimum mean-square error (MMSE) estimators, and then use the resulting channel estimates to construct the MRT precoder. In addition, average power allocation is applied in all schemes.  
Fig. \ref{fig:sumrate} illustrates the sum-rate performance when the channel strength threshold is $\beta_{\tau} = {\sigma^2}/{40}$. It is observed that the TA-based method consistently outperforms the benchmark schemes as the number of UEs increases. In the low-UE regime (e.g., $50$ to $100$ UEs), the UE-centric selection attains the slightly lower performance compared with the proposed method. Since after applying the channel strength threshold, only a small subset of UEs remains eligible for service and the system is not yet strongly interference-limited. In such conditions, the additional coordination offered by our solution is not fully exploited.

{
As mentioned earlier, Fig. \ref{fig:sumrate} compares the communication performance with and without pilot contamination caused by the charging UEs. The small performance gap between the two cases indicates that charging-UE pilot contamination has only a limited impact on communication performance, validating the corresponding simplification adopted in the analysis. }
As the number of UEs increases, the LAWN gradually enters an interference-limited and resource-constrained regime, where the structural advantage of the TA-based scheme becomes evident. Unlike the UE-centric selection, which resolves conflicts only from the UE side, the proposed method jointly accounts for both UE- and AP-side decisions. By selectively deactivating APs to mitigate interference, the TA-framework effectively reshapes the network interference structure. This joint consideration enlarges the feasible UE activation set, thereby yielding a significant sum-rate gain in overloaded scenarios. In addition, the performance gap between MMSE and LS estimation under the proposed method is marginal. In particular, in the low-UE regime, the two estimators exhibit nearly identical performance. This suggests that the dominant interference components have already been suppressed by the scheduling and activation mechanism. Consequently, the overall performance becomes less sensitive to channel-estimation accuracy, indicating that the TA framework attains interference mitigation primarily through structural decisions rather than relying on advanced signal processing.

\vspace{-0.5cm}

\subsection{Sensing Performance}

\begin{figure}
    \centering   
    \includegraphics[width=0.7\linewidth]{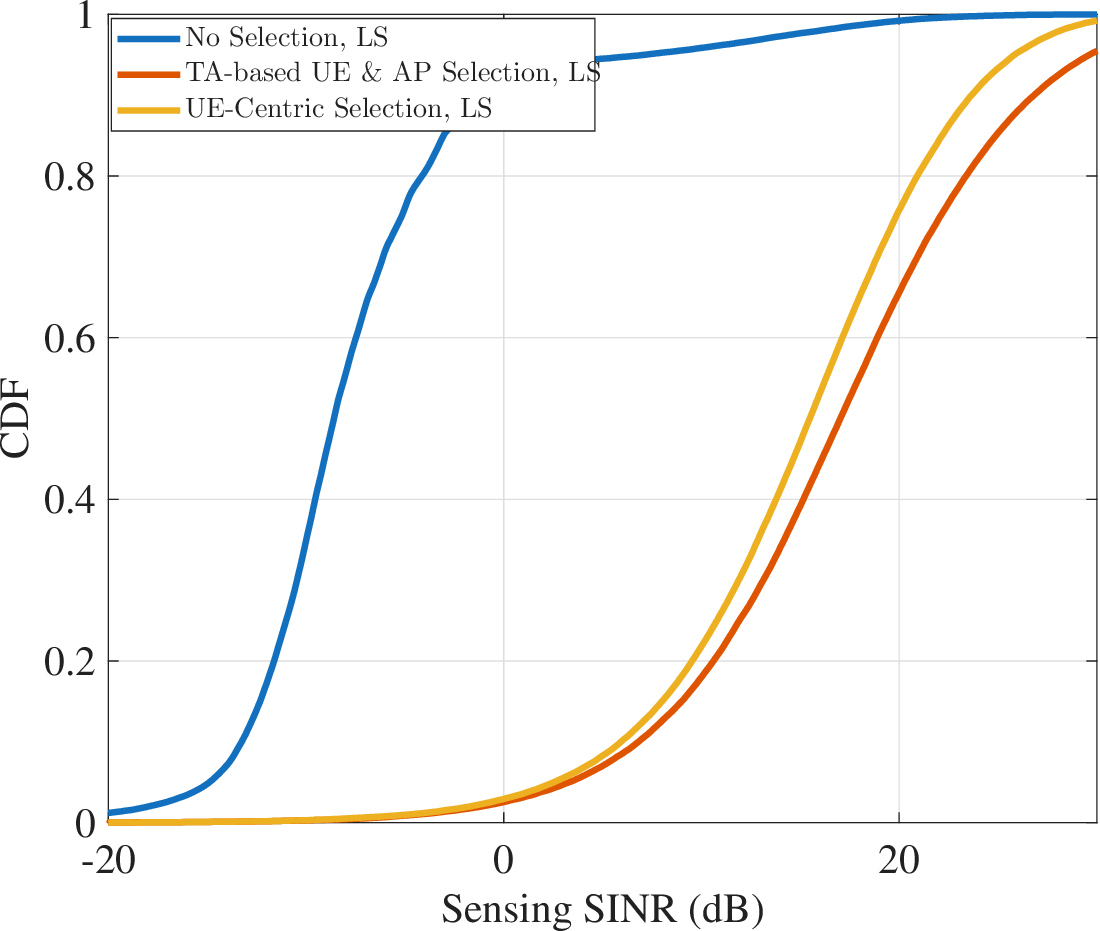}
    \caption{Localizability of target UE in the LAWN. }
    \label{fig:sensingSINR}
    \vspace{-0.5cm}
\end{figure}

{The sensing SINR is commonly adopted to determine whether a target can be detected during operation \cite{he-rss, he2025RSS}. Thus, we evaluate the sensing SINR of the target UE. Fig. \ref{fig:sensingSINR} illustrates the probability that the target UE achieves a sensing SINR above the required threshold for the proposed scheme and the benchmark methods, i.e, the detection probability of the target.} Since the performance gain of MMSE over LS estimation is negligible, we omit the MMSE-based curves in the sensing and charging evaluations for clarity. As shown in Fig. \ref{fig:sensingSINR}, the TA-based method consistently outperforms the UE-centric selection scheme. This performance gain stems from the different interference-regulation mechanisms of the two approaches. Specifically, in the UE-centric selection, dominant APs are identified primarily from the UE side; consequently, some selected “dominant” links may induce only mild interference because their channels are not sufficiently strong. In contrast, the TA-based method identifies dominant links from the AP side, facilitating the system to directly pinpoint and suppress the links that cause significant interference to the sensing target. 

\vspace{-0.5cm}

\subsection{Power Transfer Performance}

In this subsection, we evaluate the charging performance from two perspectives: (i) successful charging rate and (ii) total received energy. In the LAWN, a charging UE may harvest a non-negligible amount of energy while still failing to meet the activation threshold, i.e., $E_{k_p} < P_{\min}$. Hence, a large total received energy does not necessarily translate into a high charging success probability. Hence, we define the successful charging rate as the fraction of selected charging UEs whose received energy exceeds the required threshold. In the proposed scheme, dedicated precoders are not assigned to UEs that cannot satisfy the charging requirement, i.e., $\vv_{p,k_p}=\zero_M$; instead, transmission resources are concentrated on the UEs that are capable of being successfully charged.

\begin{figure}
    \centering   
    \includegraphics[width=0.7\linewidth]{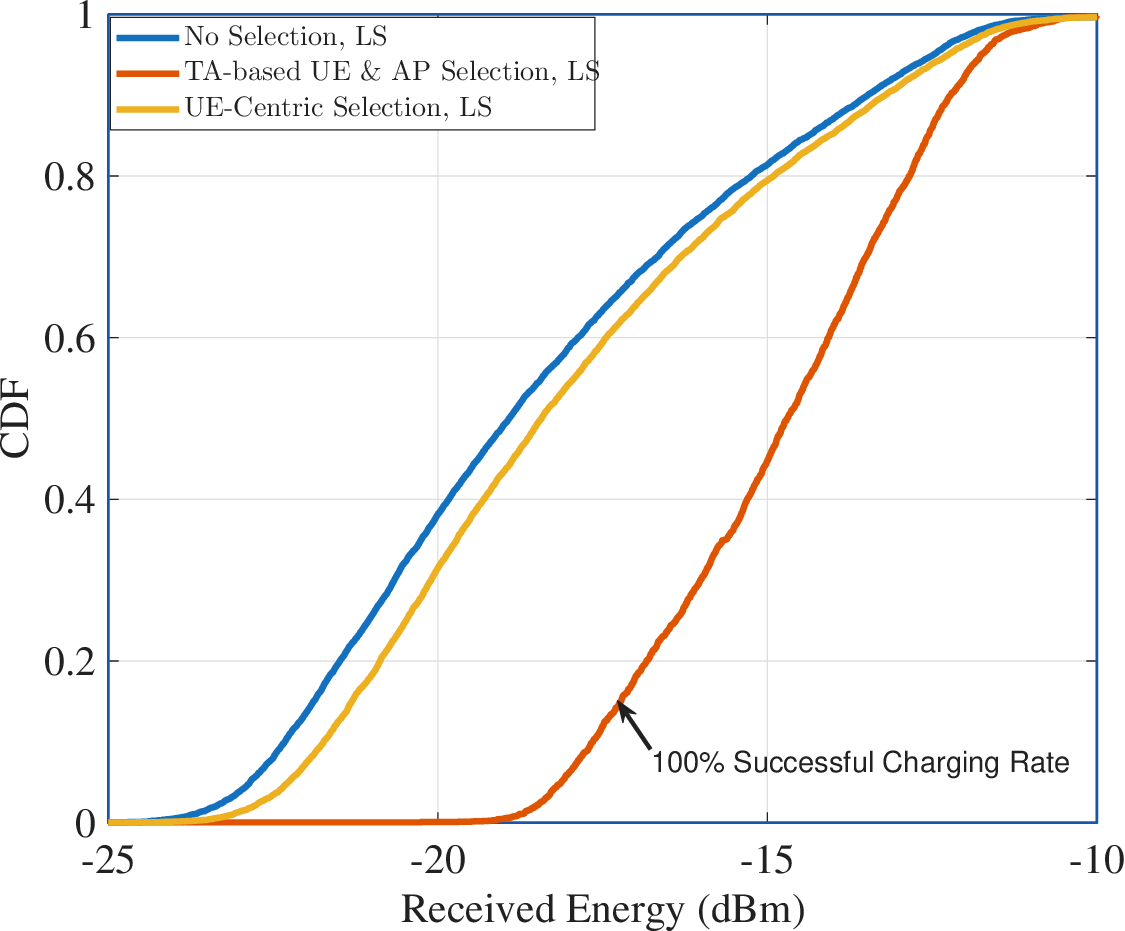}
    \caption{Successful charging rate versus the received energy}
    \label{fig:succRate}
\end{figure}

\begin{figure}
    \centering   
    \includegraphics[width=0.7\linewidth]{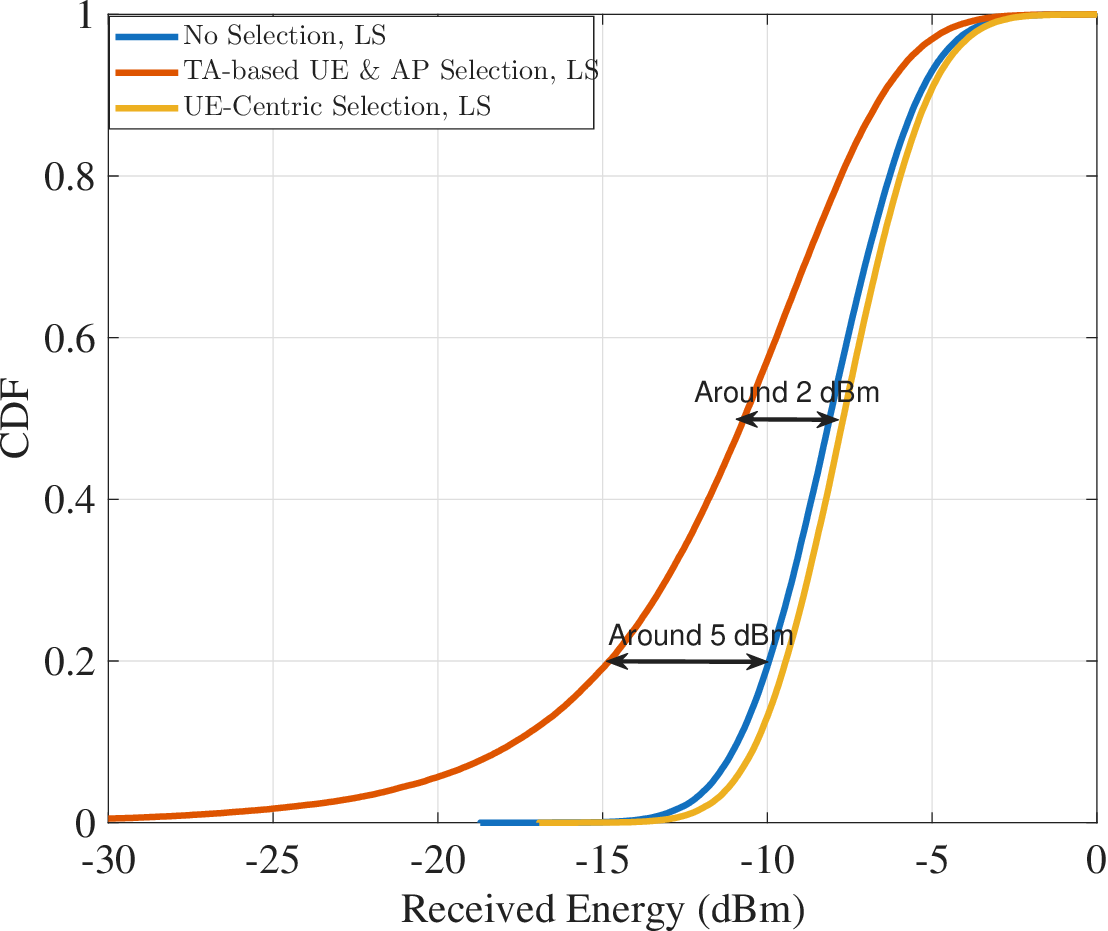}
    \caption{CDF plot of the total received energy. }
    \label{fig:SumEnergy}
    \vspace{-0.6cm}
\end{figure}

The successful charging rate and the total received energy are shown in Figs. \ref{fig:succRate} and \ref{fig:SumEnergy}, respectively. It is observed in Fig. \ref{fig:succRate} that the TA-based method achieves a $100\%$ charging success rate, significantly outperforming the user-centric and \emph{No Selection} schemes. Although our method prioritizes charging reliability, the total received energy does not degrade substantially relative to the benchmarks, as shown in Fig. \ref{fig:SumEnergy}. In most percentile ranges, the performance gap is within approximately $2$ to $3$ dBm, with only minor deviation in the lower-percentile region. These results indicate that the devised framework does not merely maximize the aggregate received energy; instead, it improves effective charging reliability through interference regulation and selective resource allocation. Consequently, the proposed methodology achieves a balanced trade-off among communication, sensing, and charging objectives.
\vspace{-0.5cm}
{
\subsection{Threshold Sensitivity Analysis}
\begin{figure}
    \centering
    \includegraphics[width=0.78\linewidth]{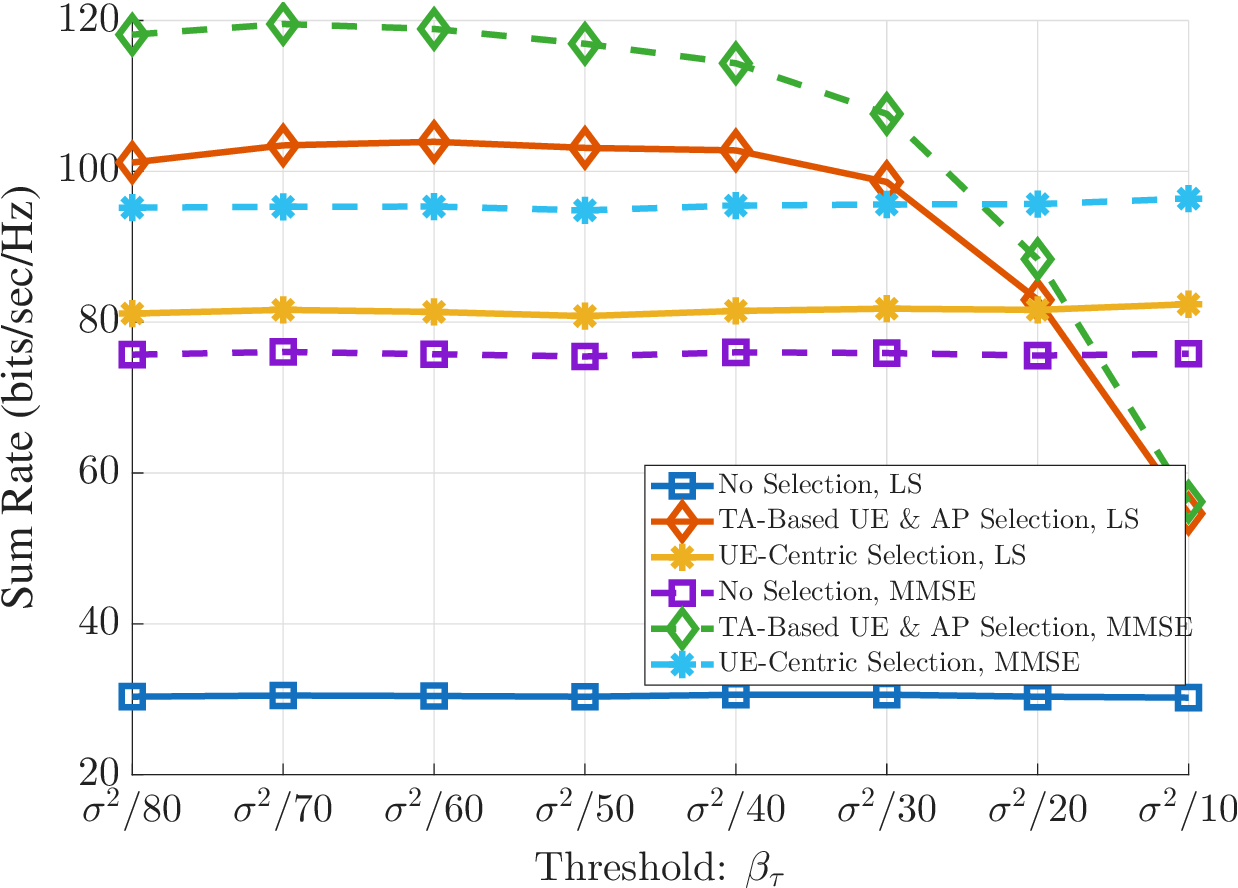}
    \caption{Sum rate versus threshold $\beta_\tau$ under 200 UEs. }
    \label{fig:beta}
\end{figure}
\begin{figure}
    \centering
    \includegraphics[width=0.75\linewidth]{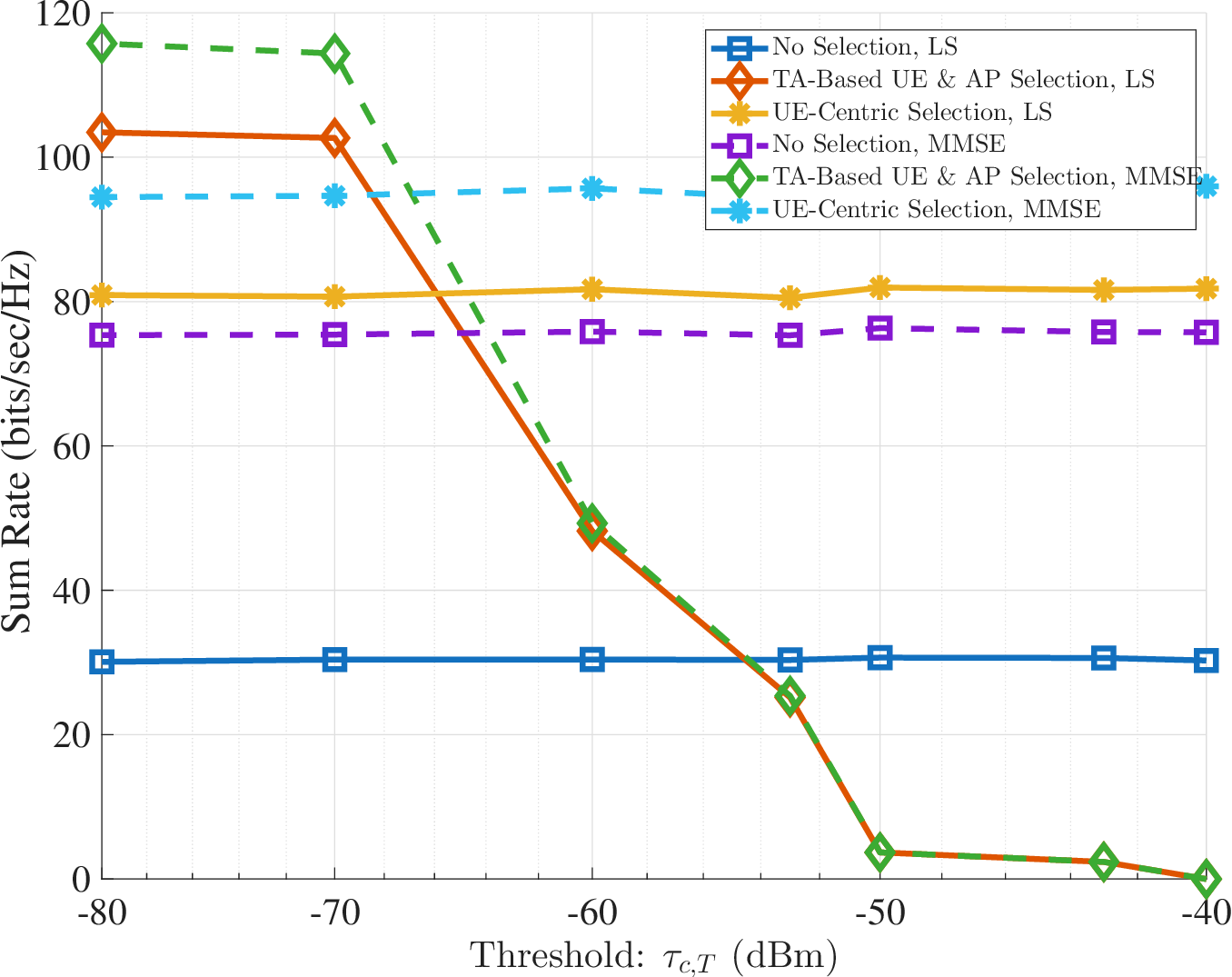}
    \caption{Sum rate versus the threshold $\tau_{c,\text{T}}$ under 200 UEs}
    \label{fig:delta}
    \vspace{-0.5cm}
\end{figure}
In the proposed method, multiple thresholds are introduced to control UE filtering and the allowable level of channel overlap. 
To investigate the effect with respect to the threshold parameters, Figs. \ref{fig:beta} and \ref{fig:delta} present the sum-rate performance versus the two channel-strength thresholds, $\beta_\tau,\tau_{c,\text{T}}$, respectively.
As discussed previously, $\beta_\tau$ is defined relative to the background-noise power $\sigma^2$ and is used to exclude UEs with excessively weak channels before executing the selection algorithm. We see in Fig. \ref{fig:beta} that when a relatively large threshold, such as $\beta_\tau=\frac{\sigma^2}{10}$, is adopted, the proposed method achieves a lower sum rate than the UE-centric selection benchmark. This is because an overly restrictive threshold classifies too many UEs as weak and significantly reduces the number of candidate UEs. As $\beta_\tau$ decreases, more potentially useful UEs are retained, and our solution begins to outperform the considered benchmarks. In particular, the performance becomes much less sensitive to $\beta_\tau$ when it is below approximately $\frac{\sigma^2}{30}$ for the LS scheme and $\frac{\sigma^2}{40}$ for the MMSE scheme. Further reductions in $\beta_\tau$ result in only minor performance variations. These results indicate that $\beta_\tau$ mainly serves as a preliminary filtering threshold that removes clearly weak UEs and reduces the computational burden, while the proposed method remains robust over a relatively broad range of sufficiently small threshold values.

A similar trend can be observed in Fig. \ref{fig:delta}. The threshold $\tau_{c,\text{T}}$ is employed within the proposed algorithm to remove relatively weak communication UEs from the current selection. Unlike the UEs filtered out by $\beta_\tau$, these UEs are not necessarily inherently weak. They may be deactivated because of the joint UE and AP selection process, including the deactivation of their serving APs, and may therefore be scheduled in subsequent time slots. When $\tau_{c,\text{T}}$ is relatively large, only a small number of communication UEs are retained, leading to lower sum rate performance than the benchmarks for both the LS and MMSE schemes. However, when $\tau_{c,\text{T}}$ is lower than approximately $-70$ dBm, the achieved sum rate remains nearly unchanged. Therefore, Figs. \ref{fig:beta} and \ref{fig:delta} demonstrate that the two thresholds effectively remove weak or relatively unfavorable UEs and reduce the computational complexity, while our approach is not overly sensitive to their exact values once they are selected within appropriate ranges. 

Another important parameter is $N_{AP,in}$, which is an integer threshold that limits the maximum number of communication and sensing connections allowed at each AP. Since UEs connected to the same AP have overlapping dominant channel supports, $N_{AP,in}$ can also be interpreted as controlling the allowable level of channel correlation among the selected UEs. For example, $N_{AP,in}=1$ requires each AP to serve at most one communication UE or sensing target. This constraint favors the selection of UEs with limited channel overlap and low mutual channel correlation, thereby yielding approximately orthogonal serving relationships. In contrast, $N_{AP,in}\geq 2$ facilitates multiple UEs or sensing targets to share the same AP and therefore permits a higher channel overlap level and correlation among the selected entities. According to the simulation results, the best performance is achieved when $N_{AP,in}=1$, whereas the sum rate decreases as $N_{AP,in}$
 increases. In the considered cell-free scenario, allowing multiple strongly connected UEs to share the same AP increases the  interference. Although a larger $N_{AP,in}$ enables more UEs to be served simultaneously, the interference increase outweighs the gain obtained from serving additional UEs.
}
\vspace{-0.3cm}
\section{Conclusion}
\label{sec:conclusion}
In this paper, we propose a TA-based optimization framework to support the efficient and reliable execution of multiple functions, including communication, sensing, and power transfer, for LAWNs. Instead of formulating the multi-objective design as a conventional optimization problem, we construct a bipartite graph based on the network topology and channel conditions. Building on this graph representation, we derive a low-complexity algorithm to efficiently handle the resulting multi-objective optimization. Simulation results show that the TA-based framework consistently outperforms typical benchmark schemes in terms of sum rate as well as successful sensing and charging probabilities, and it remains advantageous as the number of participating mobile users increases.

\bibliography{Reference_LAWN}
\bibliographystyle{IEEEtran}
\end{document}